# Geometrical Invariants of Matter Motion in Physics

Xiao Jianhua

Natural Science Foundation Research Group, Shanghai Jiaotong University, Shanghai, P.R.C.

This paper defines the spacetime geometry attached with observor as vacuum geometry (it defines the idea physical measurement geometry) and the spacetime geometry attached with matter as spacetime geometry. The initial spacetime geometry attached with matter is taken as refference geometry (named as initial co-moving coordinator system), and the current spacetime geometry attached with matter (named as current co-moving coordinator system) is determined by the four-dimensional displacement field measured in initial spacetime geometry. Matter motion is expressed by the motion transformation of basic vectors of four-dimensional co-moving coordinator system. The transformation of motion expressed by displacement describes the transient geometry form of matter motion. In general relativity theory and in gauge field theory, the geometrical invariant of "word-line distance" is used to define the geometry of space-time continuum. However, based on this research, this is only true when the time-shift is in special form. For more complicated matter motion, the gauge field theory and general relativity will be failed. The paper shows that traditional physical conservation laws and the geometrical invariant introduced by the gauge field theory and general relativity are included in the finite geometrical field theory as simple special cases.

## 1. Introduction

Hermann Weyl[1] reformulated Einstein's theory on the basis of a "pure infinitesimal geometry" into a geometrical field. In his theory, direct comparisons of length or duration could be made at neighboring points of spacetime, but not, as the Riemannian geometry of Einstein's theory permitted, "at a distance". For the continuum of spacetime, Arthur Stanley Eddington further generalized [2] Weyl's four-dimensional geometry, wherein the sole primitive geometrical notion is the non-metrical comparison of direction or orientation at the same or neighboring points.

In Weyl's geometry the magnitude of vectors at the same point, but pointing in different directions, might be directly compared to one another; in Eddington's, comparison was immediate only for vectors pointing in the same direction. His "theory of the affine field" included both Weyl's geometry and the semi-Riemannian geometry of Einstein's general relativity as special cases.

Eddington was persuaded that Weyl's "principle of relativity of length" was "an essential part of the relativistic conception". When existing physics, in particular, Einstein's theory of gravitation, is set in the context of Eddington's world geometry[3], it yields a surprising consequence: The Einstein law of gravitation appears as a definition! In the form $R_{\mu\nu} = 0$ it defines what in the "world geometry" appears to the mind as "vacuum" while in the form of the Einstein field equation noted above, it defines what is there encountered by the mind as "matter".

Based on Chen Zhida's research about the geometry of matter motion in three-dimensional space, the local geometry of matter is determined by matter motion itself rather than mathematic transformation [4-5]. So,

Weyl and Eddinton's matter geometry is damaged by confusing the phisical motion with mathematic transformation.

The paper accepts Weyl's "principle of relativity of length" as "an essential part of the relativistic conception" and Chen's idear that the locally scale (gauge) is defined by the displacement field of matter motion.

This paper reformulats the physics on the basis of displacement field of matter motion in four-dimensional geometry. The continuum of spacetime is composed by the "world geometry" defined by "vacuum" and the metter in consideration. Hence, the "matter" geometry is determined by matter motion itself rather than the "world geometry". To this purpose a general method of deductive presentation of field physics is developed mathematically by the displacement field of matter motion in the continuum of spacetime.

## 2. Spacetime Feature of Matter and Motion Transformation

The space feature of matter is that any physical matter has a definite configuration or taking up a definite space range, which can be described by three-dimensional geometry. The duration feature of matter motion is described by time in traditional physics, which is flowing with constant speed. The establishment of relativity theory shows that the time duration (in timespace attached with "observer") depends on matter motion. Hence, the "world time" invariant is introduced to define the Riemannian geometry of Einstein's theory.

The logic problem in the difinition of "world time" invariant is that it is defined as attached with "observor" (as a physical condition), while the Riemannian geometry is defined as attached with matter. Hence the time gauge in the Riemannian geometry attached with matter is variable. If the moving matter is rigid, that is the space gauges in the Riemannian geometry attached with matter is invariable, then the requiment of "world time" invariant only can be achieved by the variable time gauge attached with matter. On the other hand, if the matter is highly deformable, then the requiment of "world time" invariant will require that the time gauge attached with matter is variable. So, the difinition of "world time" invariant cannot be used as the intrinsic feature of the Riemannian geometry attached with matter.

It is accepted widely that the "world time" invariant is surportted by Michelson-Morley's experiement. But, the only conclusion one can make from the experiememnt is that the time duration defined as attached with "matter" is invariant. Therefor the the Riemannian geometry attached with matter cannot be defined by introducing the "world time" invariant.

This problem can be overcomed by introducing "world time" as time coordinator parameter in the Riemannian geometry attached with matter (be defined as four-dimensional co-moving coordinator system). By this way, a matter point is identified geometrically by its intrinsic four-dimensional position. Its motion will determine its actual gauge field which is measured by "oberservor" in conventional geometry. Hence, the matter motion will be expressed by the variation of actual gauge field. As the variation of gauge field can be expressed by the transformation of basic vectors of four-dimensional co-moving coordinator system, the motion transformation is defined to describe the matter motion.

The Riemannian geometry attached with matter in three-dimensional co-moving coordinator system has been established by Chen Zhida [4-5]. His research shows that the Riemannian geometry attached with matter is determined by the displacement field of matter, which is measured by "observor" in conventional geometry.

This paper defines the spacetime geometry attached with "observor" as "vacuum" geometry and the spacetime geometry attached with matter as the Riemannian geometry. The initial spacetime geometry attached with matter is taken as refference geometry (named as initial co-moving coordinator system), and the current spacetime geometry attached with matter (named as current co-moving coordinate system) is determined by the four-dimensional displacement field measured in initial spacetime geometry.

In the initial co-moving coordinator system, matter is identified by four-dimensional co-moving coordinators. For the matter, the four-dimensional co-moving coordinators are invariant while the actual gauge field is varying with matter motion (hence the four-dimensional co-moving coordinator system is varying with matter motion).

The initial four-dimensional co-moving coordinator system is defined by anti-covariant coordinators $(x^1, x^2, x^3, x^4)$ and initial basic vectors $(\vec{g}_1^0, \vec{g}_2^0, \vec{g}_3^0, \vec{g}_4^0)$. Where, $x^4$ is taken as time coordinator.

From physical consideration, the initial co-moving coordinator system can be defined as the standard three-dimensional space adding time-dimension, in which following conditions are met:

$$\vec{g}_i^0 \cdot \vec{g}_j^0 = \begin{cases} 0, i \neq j \\ 1, i = j \neq 4 \\ c^2, i = j = 4 \end{cases} \quad (1)$$

where, $c$ is light speed in "vacuum". The difference with the Riemannian geometry of Einstein's theory is that time gauge in here is real rather than imaginary.

By such a sellection, matter is defined in standard physical measuring system at initial time of motion.

In physics, matter motion can expressed by displacement field $u^i$ measured in standard physical measuring system. By above sellection, the displacement field $u^i$ is also defined in the initial four-dimensional co-moving coordinator system.

For any measuring time, the matter in discussing has invariant coordinators, but its local basic vectors are changed into $(\vec{g}_1, \vec{g}_2, \vec{g}_3, \vec{g}_4)$. According to Chen Zhida's research [4-5], for incremental motion of matter, there are relationship between initial basic vectors and current basic vectors:

$$\vec{g}_i = F_i^j \vec{g}_j^0 \quad (2)$$

and the transformation tensor $F_j^i$ is determined by equation:

$$F_i^j = u^j \big|_i + \delta_i^j \quad (3)$$

where, $\big|_i$ represents covariant differential, $\delta_i^j$ is unit tensor.

It is clear that the incremental motion of matter is viewed as "deformation" of matter's geometry. So, gravitation field is viewed as matter "deformation" respect with idear vacuum matter. According to Einstein's idear [6-7], the Newton's inertial coordinator system can be replaced by the gradient field of infinitisimal displacement. The equations (2) and (3) show that the displacement gradient determins the current local geometry. Hence, the transformation tensor $F_j^i$ can be named as motion transformation.

Matter motion in four-dimensional form is:

$$\begin{vmatrix} \vec{g}_1 \\ \vec{g}_2 \\ \vec{g}_3 \\ \vec{g}_4 \end{vmatrix} = \begin{vmatrix} 1+u^1\big|_1 & u^2\big|_1 & u^3\big|_1 & u^4\big|_1 \\ u^1\big|_2 & 1+u^2\big|_2 & u^3\big|_2 & u^4\big|_2 \\ u^1\big|_3 & u^2\big|_3 & 1+u^3\big|_3 & u^4\big|_3 \\ u^1\big|_4 & u^2\big|_4 & u^3\big|_4 & 1+u^4\big|_4 \end{vmatrix} \cdot \begin{vmatrix} \vec{g}_1^{\,0} \\ \vec{g}_2^{\,0} \\ \vec{g}_3^{\,0} \\ \vec{g}_4^{\,0} \end{vmatrix} \qquad (4)$$

For moving matter, its current local time gauge is:

$$\vec{g}_4 = u^1\big|_4 \cdot \vec{g}_1^{\,0} + u^2\big|_4 \cdot \vec{g}_2^{\,0} + u^3\big|_4 \cdot \vec{g}_3^{\,0} + (1+u^4\big|_4) \cdot \vec{g}_4^{\,0} \qquad (5)$$

It is clear that local time (attached with matter) measured in standard physical measuring system depends on velocity and time shift rate. Based on definition (1), the local time gauge is:

$$g_{44} = V^2 + (1+u^4\big|_4)^2 \cdot c^2 \qquad (6)$$

For observer, the standard physical measuring system is established in vaccum. If the vaccum is taken as a special matter which depends on cosmic feature matter moving in, the vaccum can be viewed as refference matter for matter motion. By this understanding, for matter moving in another refference matter field, the initial four-dimensional co-moving coordinator system can be defined by the geometry of refference matter field. In fact, for incremental motion, the refference matter field is the matter itself in initial state.

As the local time gauge in initial matter is:

$$g_{44}^{0} = c^2 \qquad (7)$$

The physical time incrementals for matter motion in initial and curent co-moving coordinator systems are, respectively:

$$dt_0 = \sqrt{g_{44}^{0}} \cdot dx^4 = c \cdot dx^4 \qquad (8)$$

$$dt = \sqrt{g_{44}} \cdot dx^4 = c \cdot \sqrt{(1+u^4\big|_4)^2 + \frac{V^2}{c^2}} \cdot dx^4 \qquad (9)$$

So, the physical time ratio between initial matter and curent moving matter is:

$$\gamma = \frac{dt_0}{dt} = \frac{\sqrt{g_{44}^{0}}}{\sqrt{g_{44}}} = \frac{1}{\sqrt{(1+u^4\big|_4)^2 + \frac{V^2}{c^2}}} \qquad (10)$$

If:

$$\big|u^4\big|_4\big| << 1 \qquad (11)$$

then, for low velocity motion of matter,

$$\gamma_{low} = \frac{dt_0}{dt} = \frac{1}{\sqrt{1+\frac{V^2}{c^2}}} \approx \sqrt{1-\frac{V^2}{c^2}} \qquad (12)$$

It can be seen that the time incremental of moving matter is longer than the time incremental of static matter. Noted that our geometry is defined in commoving-coordinator system, for such a selection, $dt$ is static time while the $dt_0$ is the moving time in general relativity. So, in fact this result is the same as the result of special relativity.

For high speed moving, unlike the special relativity, the physical time shorting will cause negative time displacement gradient. There are two special cases.

(a). For very high speed motion, if:

$$(1+\frac{\partial u^4}{\partial x^4})^2 << \frac{V^2}{c^2} \tag{13}$$

then:

$$\gamma_{high} = \frac{1}{\sqrt{(1+\frac{\partial u^4}{\partial x^4})^2 + \frac{V^2}{c^2}}} \approx \frac{c}{V} \tag{14}$$

It shows that for very high speed moving the physical time ratio will become bigger than one rather than less than one for the low speed moving. This overcomes the twin-brother's porodox in the Riemannian geometry of Einstein's theory.

(b). There exists such a matter motion, the matter moving speed is very low or zero, but there is significant time displacement gradient that:

$$(1+\frac{\partial u^4}{\partial x^4})^2 >> \frac{V^2}{c^2} \tag{15}$$

For such a matter motion, the physical time ratio is:

$$\gamma_{timeshift} \approx \frac{1}{1+\frac{\partial u^4}{\partial x^4}} \approx 1 - \frac{\partial u^4}{\partial x^4} \tag{16}$$

It may be less or bigger than one, depending on the sign of time displacement gradient.

As the time shorting and time expanding are both well observed in cosmic oberservation [8], and the general relativity has confirmed that the oberserved physical time is a relative one and is variable, it is meaningful to introduce such a motion matter.

## 3. Geometrical Invariants and Conservation Laws in Physics

For a matter, if it is defined initially in vacuum geometry as having geometric invariant:

$$ds_0^2 = g_{11}^0(dx^1)^2 + g_{22}^0(dx^2)^2 + g_{33}^0(dx^3)^2 + c^2(dx^4)^2 \tag{17}$$

Then, the matter in physical motion will have current geometric invariant as:

$$ds^2 = [g_{11}^0 + (c\frac{\partial u^4}{\partial x^1})^2](dx^1)^2 + [g_{22}^0 + (c\frac{\partial u^4}{\partial x^2})^2](dx^2)^2 + [g_{33}^0 + (c\frac{\partial u^4}{\partial x^3})^2](dx^3)^2$$

$$+ 2c^2\frac{\partial u^4}{\partial x^1}\frac{\partial u^4}{\partial x^2}dx^1dx^2 + 2c^2\frac{\partial u^4}{\partial x^2}\frac{\partial u^4}{\partial x^3}dx^2dx^3 + 2c^2\frac{\partial u^4}{\partial x^3}\frac{\partial u^4}{\partial x^1}dx^3dx^1$$

$$+[V^2 + c^2 + 2c^2 \frac{\partial u^4}{\partial x^4} + c^2(\frac{\partial u^4}{\partial x^4})^2](dx^4)^2 \tag{18}$$

Hence, we have:

$$ds^2 - ds_0^2 = (c\frac{\partial u^4}{\partial x^1})^2(dx^1)^2 + (c\frac{\partial u^4}{\partial x^2})^2(dx^2)^2 + (c\frac{\partial u^4}{\partial x^3})^2(dx^3)^2$$

$$+ 2c^2 \frac{\partial u^4}{\partial x^1}\frac{\partial u^4}{\partial x^2}dx^1 dx^2 + 2c^2 \frac{\partial u^4}{\partial x^2}\frac{\partial u^4}{\partial x^3}dx^2 dx^3 + 2c^2 \frac{\partial u^4}{\partial x^3}\frac{\partial u^4}{\partial x^1}dx^3 dx^1$$

$$+[V^2 + 2c^2 \frac{\partial u^4}{\partial x^4} + c^2(\frac{\partial u^4}{\partial x^4})^2](dx^4)^2 \tag{19}$$

So, geometric invariant for physical motion can be defined as:

$$d\Sigma^2 = ds^2 - ds_0^2 = c^2(du^4)^2 + 2c^2 \frac{\partial u^4}{\partial x^4}(dx^4)^2 + V^2(dx^4)^2 \tag{20}$$

where:

$$(du^4)^2 = (\frac{\partial u^4}{\partial x^1})^2(dx^1)^2 + (\frac{\partial u^4}{\partial x^2})^2(dx^2)^2 + (\frac{\partial u^4}{\partial x^3})^2(dx^3)^2 + (\frac{\partial u^4}{\partial x^4})^2(dx^4)^2$$

$$+ 2\frac{\partial u^4}{\partial x^1}\frac{\partial u^4}{\partial x^2}dx^1 dx^2 + 2\frac{\partial u^4}{\partial x^2}\frac{\partial u^4}{\partial x^3}dx^2 dx^3 + 2\frac{\partial u^4}{\partial x^3}\frac{\partial u^4}{\partial x^1}dx^3 dx^1 \tag{21}$$

From physical consideration, no matter what initial geometry system is used as initial co-moving coordinator system geometry, the subjectivity of physical motion will require the $d\Sigma^2$ be geometrical invariant.

## 4. Matter Motion
Matter motion can be divided into three simple typical cases, as described bellow.
### 4.1 Non-Quantum Matter Motion
Based on the previous research, for non-quantum matter motion the time displacement field component can be replaced by Newton's Acceleration Law. It is equivalent to suppose there is no time displacement. Hence, we suppose for non-quantum matter motion:

$$V^2 >> 2c^2 \frac{\partial u^4}{\partial x^4} + c^2(du^4)^2 \tag{22}$$

Hence, the geometrical invariant of non-quantum matter motion is:

$$d\Sigma^2 = V^2(dx^4)^2 \tag{23}$$

That is:

$$d\Sigma^2 = (du^1)^2 + (du^2)^2 + (du^3)^2 \tag{24}$$

This defines the conventional three-dimensional physical measure space. It defines the spatial motion distance as geometrical invariant. For Newton's matter point, it leads to the moving moment conservation.

**4.2 Quantum Matter Motion**

Based on the previous research, for quantum matter motion the time displacement field component play the main roles. It is equivalent to suppose there is a significant time displacement. Hence, we suppose for quantum matter motion:

$$V^2 + 2c^2 \frac{\partial u^4}{\partial x^4} \ll c^2 (du^4)^2 \tag{25}$$

Hence, the geometrical invariant of quantum matter motion is:

$$d\Sigma^2 = ds^2 - ds_0^2 = c^2 (du^4)^2 \tag{26}$$

So, we can define time displacement is a scalar function in four-dimensional time-space. As a special case, this geometrical invariant can be explained as matter-energy conservation.

**4.3 Very High Speed Matter Motion**

For very high speed matter motion, the time gradient of time displacement can be ignored, that is to say:

$$V^2 \gg 2c^2 \frac{\partial u^4}{\partial x^4} + c^2 (\frac{\partial u^4}{\partial x^4})^2 \tag{27}$$

Then we will have the geometrical invariant as:

$$d\Sigma^2 = ds^2 - ds_0^2 = c^2 (d\tilde{u}^4)^2 + (du^1)^2 + (du^2)^2 + (du^3)^2 \tag{28}$$

where:

$$(d\tilde{u}^4)^2 = (\frac{\partial u^4}{\partial x^1})^2 (dx^1)^2 + (\frac{\partial u^4}{\partial x^2})^2 (dx^2)^2 + (\frac{\partial u^4}{\partial x^3})^2 (dx^3)^2$$

$$+ 2\frac{\partial u^4}{\partial x^1}\frac{\partial u^4}{\partial x^2} dx^1 dx^2 + 2\frac{\partial u^4}{\partial x^2}\frac{\partial u^4}{\partial x^3} dx^2 dx^3 + 2\frac{\partial u^4}{\partial x^3}\frac{\partial u^4}{\partial x^1} dx^3 dx^1 \tag{29}$$

It can be seen that $d\tilde{u}^4$ is independent with $dx^4$, so it is a scalar defined in three-dimensional space.

For matter behaves as the time displacement defined by the following wave form:

$$u^4 = U_0 \exp[(\vec{R} \cdot \vec{k} \pm \frac{c}{n} t) j] \tag{30}$$

or matter behaves as the time displacement defined by the following spatial-localized harmonic particle form:

$$u^4 = U_0 \exp(\vec{R} \cdot \vec{l} \pm \frac{c}{n} tj) \tag{31}$$

If we take the definition of:

$$d\tilde{u}^4 = jd\tau \tag{32}$$

and explain the $d\tau$ as the universal time (world time), we will get the geometrical invariant as:

$$d\Sigma^2 = c^2 (d\tilde{u}^4)^2 + (du^1)^2 + (du^2)^2 + (du^3)^2$$

$$= (du^1)^2 + (du^2)^2 + (du^3)^2 - c^2 (d\tau)^2 \tag{33}$$

In general relativity theory and in gauge field theory, this equation form is used to define the geometry of space-time continuum under the meaning that the $d\Sigma^2$ is the geometrical invariant of "word-line distance". However, based on this paper's research, this is only true when the time displacement is normal and in form (30) or (31). For more complicated matter motion, the gauge field theory and general relativity will be failed.

Summing up above discussion, we can see that traditional physical conservation laws and the geometrical invariant introduced by the gauge field theory and general relativity are included in the finite geometrical field theory as simple special cases.

## 5. Conclusion

The research shows an unified field theory of physics can be established with the idea introducing motion transformation, which is completely determined by the displacement gradient in four-dimensional space-time continuum. The vacuum is taken as a special matter with one intrinsic physical parameter-light speed $c$. For a matter, it can be identified with co-moving coordinators with local varying geometry. The local varying geometry represents the motion of matter in consideration. The paper introduces the geometrical invariants to express physical conservation laws. The results show that the geometrical invariant introduced by the gauge field theory and general relativity are included in the finite geometrical field theory as simple special cases.

So, an unified physics of matter motion can be expressed by the finite geometrical field theory given by this paper.

**Refference**
[1] Eddington, A.S., A Generalization of Weyl's Theory of the Electromagnetic and Gravitational Fields, *Proceedings of the Royal Society of London,* 1921, *Series A* 99, 104-122.
[2] Eddington, A.S., *The Mathematical Theory of Relativity,* Cambridge: Cambridge University Press. 1923.
[3] Eddington, A.S., *Fundamental Theory.* Cambridge: Cambridge University Press. Posthumously published. 1948.
[4] Chen Zhida., Geometric Theory of Finite Deformation Mechanics for Continuum. *Acta Mechanica Sinica,* 1979, No.2, 107-117 (In Chinese).
[5] Chen Zhida., *Rational Mechanics.* Chongqing: Chongqing Publication, 2000 (In Chinese).
[6] Einstein, A., *Relativistic Theory of Non-Symmetric Field,* In Einstein's Work Collection (Vol.2), Commercial Pub., 1977, p560-565 (in Chinese). (Original: Einstein, A., Meaning of Relativity, (5[th] edition), 1954, p133-166).
[7] Einstein, A., *Foundation of General Relativity,* In Einstein's Work Collection (Vol.2), Commercial Pub., 1977, p292 (in Chinese). (Original: Einstein, A.,Die Grundlage der Allgemeinen Relativitätstheorie, Annalen der Physik, V4-49, 1916, p769-822).
[8] He Xiantau, *Observation Cosmics*, Beijing: Scientific Publication, 2002 (In Chinese).